# Title: Discovery of a Cooper-Pair Density Wave State in a Transition-Metal Dichalcogenide

**Authors:**

Xiaolong Liu[1§], Yi Xue Chong[1§], Rahul Sharma[1,2], and J.C. Séamus Davis[1,3,4,5]

**Affiliations:**

1. *LASSP, Department of Physics, Cornell University, Ithaca NY 14850, USA*
2. *Department of Physics, University of Maryland, College Park MD 20740, USA*
3. *Department of Physics, University College Cork, Cork T12 R5C, IE*
4. *Max-Planck Institute for Chemical Physics of Solids, D-01187 Dresden, DE*
5. *Clarendon Laboratory, University of Oxford, Oxford, OX1 3PU, UK*

§ *These authors contributed equally to this project.*

**Abstract:**

Pair density wave (PDW) states are defined by a spatially modulating superconductive order-parameter. To search for such states in transition metal dichalcogenides (TMD) we use high-speed atomic-resolution scanned Josephson-tunneling microscopy (SJTM). We detect a PDW state whose electron-pair density and energy-gap modulate spatially at the wavevectors of the preexisting charge density wave (CDW) state. The PDW couples linearly to both the *s*-wave superconductor and to the CDW, and exhibits commensurate domains with discommensuration phase-slips at the boundaries, conforming to those of the lattice-locked commensurate CDW. Nevertheless, we find a global $\delta\Phi \cong \pm 2\pi/3$ phase difference between the PDW and CDW states, possibly owing to the Cooper-pair wavefunction orbital content. Our findings presage pervasive PDW physics in the many other TMDs that sustain both CDW and superconducting states.

**Main Text:**

**1** Transition metal dichalcogenides (TMDs) are a rich platform for the exploration of quantum matter (*1*-*6*). In this context, a fundamental but elusive state is the electron-pair density wave (PDW) (*7*). Theoretically, the PDW state of TMDs was predicted to be generated by magnetic-field (*8*), by spin-valley locking (*9*), by a CDW lock-in transition (*10*), and by doping TMD bilayers (*11*); PDW melting into a charge-6e superfluid was also predicted (*12*). However, detecting the PDW state in a TMD experimentally has been challenging.

**2** A familiar TMD state is the charge density wave (CDW): a charge density field $\rho_{\mathrm{C}}(\boldsymbol{r})$ that modulates spatially at wavevectors $\boldsymbol{Q}_{\mathrm{C}}^{i}$ $(i = 1, 2, 3)$ which are not crystal reciprocal lattice vectors. Such a charge density modulation

$$\rho_{\mathrm{C}}^{i}(\boldsymbol{r}) = \rho_i(\boldsymbol{r}) e^{i\boldsymbol{Q}_{\mathrm{C}}^{i}\cdot\boldsymbol{r}} + \rho_i^*(\boldsymbol{r}) e^{-i\boldsymbol{Q}_{\mathrm{C}}^{i}\cdot\boldsymbol{r}} \tag{1}$$

has a mean-field order parameter $\langle c_{\boldsymbol{k}}^{\dagger} c_{\boldsymbol{k}+\boldsymbol{Q}_{\mathrm{C}}^{i}} \rangle$, where $c_{\boldsymbol{k}}^{\dagger}(c_{\boldsymbol{k}})$ are the creation (annihilation) operators for single-electron $\boldsymbol{k}$-space eigenstates. The simplest TMD superconductor state is spatially homogeneous but breaks gauge symmetry

$$\Delta_{\mathrm{S}}(\boldsymbol{r}) = \Delta_0 e^{i\phi} \tag{2}$$

with a mean-field order parameter $\langle c_{\boldsymbol{k}}^{\dagger}\, c_{-\boldsymbol{k}}^{\dagger} \rangle$ . By contrast, a PDW state is described by an electron-pair field $\Delta_{\mathrm{P}}(\boldsymbol{r})$ that modulates spatially at one or more wavevectors $\boldsymbol{Q}_{\mathrm{P}}^{i}$

$$\Delta_{\mathrm{P}}^{i}(\boldsymbol{r}) = \left[\Delta_i(\boldsymbol{r}) e^{i\boldsymbol{Q}_{\mathrm{P}}^{i}\cdot\boldsymbol{r}} + \Delta_i^*(\boldsymbol{r}) e^{-i\boldsymbol{Q}_{\mathrm{P}}^{i}\cdot\boldsymbol{r}}\right] e^{i\theta} \tag{3}$$

This state also breaks gauge symmetry and its mean-field order parameter is $\langle c_{\boldsymbol{k}}^{\dagger}\, c_{-\boldsymbol{k}+\boldsymbol{Q}_{\mathrm{P}}^{i}}^{\dagger} \rangle$. Sophisticated atomic-scale visualization of TMD states using single-electron tunneling (*13-16*) has revealed CDW quantum phase transitions (*13*), a CDW Bragg glass (*14*), interfacial band alignment (*15*), and strain control of the CDW state(*16*). But to detect and image a PDW state in TMDs remained an experimental challenge.

**3** Experimentally, the total electron-pair density $\rho_{\mathrm{CP}}(\boldsymbol{r})$ might be visualized by measuring Josephson critical-current $I_{\mathrm{J}}(\boldsymbol{r})$ to a superconducting STM tip (*17*), because $\rho_{\mathrm{CP}}(\boldsymbol{r}) \propto I_{\mathrm{J}}^2(\boldsymbol{r}) R_{\mathrm{N}}^2(\boldsymbol{r})$ where $R_{\mathrm{N}}$ is the normal-state junction resistance (*18,19*). But this has proven impractical because the thermal fluctuation energy $k_{\mathrm{B}}T$ typically exceeds the Josephson energy $E_{\mathrm{J}} = \Phi_0 I_{\mathrm{J}}/2\pi$ where

$$I_{\mathrm{J}} = \frac{\pi\Delta(T)}{2eR_{\mathrm{N}}}\tanh\left(\frac{\Delta(T)}{2k_{\mathrm{B}}T}\right) \tag{4}$$

($k_{\mathrm{B}}$: Boltzmann constant, 2e: electron-pair charge, $\Phi_0$: magnetic flux quantum). Instead, when $E_{\mathrm{J}} < k_{\mathrm{B}}T$, the tip-sample Josephson junction exhibits a phase-diffusive (*20-22*) steady-state at voltage $V$ and electron-pair current

$$I_{\mathrm{CP}}(V) = \frac{1}{2}I_{\mathrm{J}}^2 Z\, V/(V^2 + V_{\mathrm{c}}^2) \tag{5}$$

Here $V_{\mathrm{c}} = 2eZk_{\mathrm{B}}T/\hbar$ with $Z$ the high-frequency impedance in series with the voltage source. From Eq. 5

$$dI_{\mathrm{CP}}/dV \equiv g(V) = \frac{1}{2}I_{\mathrm{J}}^2 Z\,(V_{\mathrm{c}}^2 - V^2)/(V^2 + V_{\mathrm{c}}^2)^2 \tag{6}$$

yields $g(0) \propto I_{\mathrm{J}}^2$ (Section 1 of (*23*), Fig. S1). Thus, spatially resolved measurements of $g(0,\boldsymbol{r})$ can provide a practical means (*24-28*) to image $I_{\mathrm{J}}(\boldsymbol{r})$, so that the electron-pair density can then be visualized as $N_{\mathrm{CP}}(\boldsymbol{r}) \equiv g(\boldsymbol{r},0)R_{\mathrm{N}}^2(\boldsymbol{r}) \propto \rho_{\mathrm{CP}}(\boldsymbol{r})$ (Section 1 of (*23*)).

***4*** Here we study bulk crystals of 2H-$NbSe_2$, a quasi-two-dimensional TMD with a robust CDW state (*29*). It has a hexagonal layered structure with Se-Se separation $d$ and a Fermi surface with pockets surrounding the Γ and K points (Fig. S2). The CDW phase transition at $T \approx 33.5$ K generates crystal and charge density modulations at three in-plane wavevectors $\boldsymbol{Q}_{\mathrm{C}}^i \approx \{(1,0); (1/2, \sqrt{3}/2); (-1/2, \sqrt{3}/2)\}2\pi/3a_0$ ($a_0 = \sqrt{3}d/2$ is the unit cell dimension), and the s-wave superconductivity (SSC) transition at $T_{\mathrm{C}} \approx 7.2$ K completely gaps the Fermi surface. We use atomic-resolution superconducting scan tips made of Nb (17) with standard tip-energy-gap $|\Delta_{\mathrm{T}}| \approx 0.9$ meV (Fig. S3).

**5** Figure 1A shows a typical topographic image $T(\boldsymbol{r}, V)$ of the Se-termination layer of $NbSe_2$ when using such tips, with the CDW modulations appearing as $3a_0$ periodic intensity amplifications (*13*, *14*) (inset $T(\boldsymbol{q}, V)$). Figure 1B shows a typical differential tunneling conductance spectrum $dI/dV|_V \equiv g(V)$. To simultaneously visualize the CDW, $s$-wave superconductor (SSC) and any putative PDW states, a dynamic range exceeding $10^4$ is required in the tip-sample voltage, spanning the CDW range from above ~50 mV (Fig. 1B), to the SSC energy gap range ~1 mV (Fig. 1C), to the Josephson pair-current range approaching ~10 μV (Fig. 1, D and E). Visualizing the quasiparticle densities $N_{\mathrm{Q}}(\boldsymbol{r})$ of both CDW and SSC uses single-electron tunneling at energies indicated by the red (Fig. 1B) and green arrows (Fig. 1C), respectively. Visualizing electron-pair density $N_{\mathrm{CP}}(\boldsymbol{r})$ of the condensate utilizes the phase-diffusive Josephson tunneling current $I_{\mathrm{CP}}(V)$ or $g(0)$ indicated by the blue arrows (Fig. 1, D and E).

**6** At $T = 290$ mK, we first image $N_{\mathrm{Q}}(\boldsymbol{r}) \equiv g(r, -20\text{ mV})$ at $V = -20$ mV where CDW intensity is strong (*13*), with results shown in Fig. 2A. Next, we image the normal-state resistance (Fig. S4) of the tip-sample Josephson junction $R_{\mathrm{N}}(\boldsymbol{r}) \equiv I^{-1}(\boldsymbol{r}, -4.5\text{ mV})$ as shown in Fig. 2B. Third, we study the electron-pair current by measuring $g(\boldsymbol{r}, 0)$ (Eq. 6) as shown in Fig. 2C. All four independent images $T(\boldsymbol{r}, V)$: $N_{\mathrm{Q}}(\boldsymbol{r})$: $R_{\mathrm{N}}(\boldsymbol{r})$: $g(\boldsymbol{r}, 0)$ are registered to each other with precision of $\delta x \approx \delta y \lesssim 15$ pm (Section 2 of (*23*), Fig. S5). This constitutes a typical data set for visualizing the crystal, CDW, SSC and PDW states simultaneously; its acquisition required developing high-speed scanned Josephson-tunneling microscopy (SJTM) imaging protocols Section 3 of (*23*), Fig. S6). Eventually, to visualize the electron-pair density, we use the data in Fig. 2, B and C, to derive $N_{\mathrm{CP}}(\boldsymbol{r}) \equiv g(\boldsymbol{r}, 0)R_{\mathrm{N}}^2(\boldsymbol{r})$ as shown in Fig.

2D. Here, we see electron-pair density modulations at three in-plane wavevectors $\boldsymbol{Q}_{\mathrm{P}}^{i} \approx \{(1,0); (1/2, \sqrt{3}/2); (-1/2, \sqrt{3}/2)\} 2\pi/3a_0$, indicated by the blue circles in $N_{\mathrm{CP}}(\boldsymbol{q})$ inset to Fig. 2D. Ultimately, Fig. 2E shows $N_{\mathrm{C}}(\boldsymbol{r})$ of the CDW containing only the charge density modulations at $\boldsymbol{Q}_{\mathrm{C}}^{i}$ (Fig. 2A), whereas Fig. 2F shows the simultaneous $N_{\mathrm{P}}(\boldsymbol{r})$ of the PDW containing only the electron-pair density modulations at $\boldsymbol{Q}_{\mathrm{P}}^{i}$ (Fig. 2D). The data in Fig. 2F represent the observation of a PDW state in a TMD material, $NbSe_2$.

*7* In a PDW state, the energy gap $\Delta_{\mathrm{P}}(\boldsymbol{r})$ should also modulate at $\boldsymbol{Q}_{\mathrm{P}}$ (Eq. 3). We define the total gap $|\Delta(\boldsymbol{r})| = |\Delta_{P}(\boldsymbol{r}) + \Delta_{S}|$ to be half the energy separation between two coherence peaks minus $|\Delta_{\mathrm{T}}|$ (Fig. 1C). Our measured $|\Delta(\boldsymbol{r})|$ then exhibits modulations at three wavevectors $\boldsymbol{Q}_{\mathrm{P}}^{i} \approx \{(1,0); (1/2, \sqrt{3}/2); (-1/2, \sqrt{3}/2)\} 2\pi/3a_0$ (Section 4 of (*23*) and Fig. S7). This confirms independently, using single-electron tunneling, the existence of a PDW state in $NbSe_2$. Its gap modulation amplitude $|\Delta_{\mathrm{P}}| < 0.01|\Delta_0|$ (Section 4 of (*23*)). Figure 3A shows a plot of the measured Fourier amplitudes of simultaneous $N_{\mathrm{Q}}(|\boldsymbol{q}|)$ and $N_{\mathrm{CP}}(|\boldsymbol{q}|)$ in the directions of $\boldsymbol{Q}_{\mathrm{P}}^{i} \approx \boldsymbol{Q}_{C}^{i}$. The key maxima occur near $|\boldsymbol{q}| = 2\pi/3a_0$ establishing quantitatively that $|\boldsymbol{Q}_{\mathrm{P}}^{i}| = |\boldsymbol{Q}_{\mathrm{C}}^{i}| \pm 1\%$. But, although imaged in precisely the same FOV, the charge density modulations (Fig. 2E) and electron-pair density modulations (Fig. 2F) appear distinctly different, with normalized cross correlation coefficient $\eta \approx -0.4$.

*8* Possible microscopic mechanisms for a PDW state include Zeeman splitting (*30,31*) of a Fermi surface (not relevant here), and strongly correlated electron-electron interactions generating intertwined CDW and PDW states (*32,33*). But whatever the microscopic PDW mechanism for $NbSe_2$, Ginzburg-Landau (GL) theory allows a general analysis of interactions between SSC and CDW states. Consider a didactic GL free energy density

$$\mathcal{F} = \mathcal{F}_\mathrm{S} + \mathcal{F}_\mathrm{C} + \mathcal{F}_\mathrm{P} + \sum_i (\lambda_i \rho_i \Delta_\mathrm{S}^* \Delta_i + c.c.) \quad (7)$$

Here $\mathcal{F}_\mathrm{S}, \mathcal{F}_\mathrm{C}, \mathcal{F}_\mathrm{P}$ are the free energy densities of an SSC state (Eq. 2), a CDW state (Eq. 1), and a PDW state (Eq. 3), respectively. The term $\lambda_i \rho_i \Delta_\mathrm{S}^* \Delta_i$ represents lowest order coupling of the SSC and CDW states with a PDW, and induces $\Delta_\mathrm{P}^i(\boldsymbol{r})$ at wavevectors $\boldsymbol{Q}_\mathrm{P}^i = \boldsymbol{Q}_\mathrm{C}^i$ due to interactions of $\rho_\mathrm{C}^i(\boldsymbol{r})$ and $\Delta_\mathrm{S}$. But the relative spatial arrangements of $\rho_\mathrm{C}^i(\boldsymbol{r})$ and $\rho_\mathrm{P}^i(\boldsymbol{r}) \propto |\Delta_\mathrm{P}^i(\boldsymbol{r})|^2$ are ambiguous because, if $\rho_\mathrm{C}^i(\boldsymbol{r}) \propto \cos[\boldsymbol{Q}_\mathrm{C}^i \cdot \boldsymbol{r} + \Phi_\mathrm{C}^i(\boldsymbol{r})]$ and $\rho_\mathrm{P}^i(\boldsymbol{r}) \propto \cos[\boldsymbol{Q}_\mathrm{P}^i \cdot \boldsymbol{r} + \Phi_\mathrm{P}^i(\boldsymbol{r})]$ and $\boldsymbol{Q}_\mathrm{P}^i = \boldsymbol{Q}_\mathrm{C}^i$, the phase difference $\delta\Phi^i(\boldsymbol{r}) \equiv \Phi_\mathrm{P}^i(\boldsymbol{r}) - \Phi_\mathrm{C}^i(\boldsymbol{r})$ cannot be predicted from Eq. 7.

**9** To explore the Ginzburg-Landau predictions we next visualize quasiparticle density $N_\mathrm{Q}(\boldsymbol{r})$ and electron-pair density $N_\mathrm{CP}(\boldsymbol{r})$ centered on a quantized vortex core (Section 5 of (*23*), Figs. S8, S9). $N_\mathrm{P}(\boldsymbol{r})$ of the PDW is shown in Fig. 3B and the total $N_\mathrm{CP}(\boldsymbol{r})$ is shown in Fig. 3C. The background superfluid density $N_\mathrm{S}(\boldsymbol{r}) = N_\mathrm{CP}(\boldsymbol{r}) - N_\mathrm{P}(\boldsymbol{r})$ is azimuthally symmetric about the core, consistent with previous experiments (*34*) and GL theory (*35*). The mutual decay of the PDW and SSC into the vortex along the yellow dashed lines in Fig. 3, B and C is visualized in Fig. 3, D and E. More quantitatively, if $N_\mathrm{C}^i(\boldsymbol{r}) = A_\mathrm{C}^i(\boldsymbol{r})\cos[\boldsymbol{Q}_\mathrm{C}^i \cdot \boldsymbol{r} + \Phi_\mathrm{C}^i(\boldsymbol{r})]$ and $N_\mathrm{P}^i(\boldsymbol{r}) = A_\mathrm{P}^i(\boldsymbol{r})\cos[\boldsymbol{Q}_\mathrm{P}^i \cdot \boldsymbol{r} + \Phi_\mathrm{P}^i(\boldsymbol{r})]$, the combined PDW amplitude is represented by $A_\mathrm{P}^\mathrm{RMS}(\boldsymbol{r}) = \sqrt{\sum_{i=1}^{3} [A_\mathrm{P}^i(\boldsymbol{r})]^2 / 3}$ in Fig. 3F ( Section 6 of (*23*)), demonstrating its mutual decay with $N_\mathrm{S}(\boldsymbol{r})$ of SSC in Fig. 3G. This is as expected within GL theory (Eq. 7) for a CDW-induced PDW state.

**10** Even though the PDW state is strongly linked to parent SSC state (Fig. 3) and to the modulation wavevectors of the CDW state (Figs. 2D, 3A), the two modulating states appear spatially disparate at atomic-scale (Fig. 2, E and F). To explore this unexpected phenomenon, we visualize the amplitude and phase of the CDW and PDW for all three wavevectors $\boldsymbol{Q}_{\mathrm{C}}^{i} \approx \boldsymbol{Q}_{\mathrm{P}}^{i}$ (Fig. S10). Figure 4A shows measured $A_{\mathrm{C}}^{1}(\boldsymbol{r})$ from Fig. 2A, and Fig. 4B the simultaneously measured $A_{\mathrm{P}}^{1}(\boldsymbol{r})$ from Fig. 2D. Both show nanoscale variations in the magnitude of their order parameters that are spatially alike, consistent with Eq. 7. Figures 4, C and D, are the $\Phi_{\mathrm{C}}^{1}(\boldsymbol{r})$ and $\Phi_{\mathrm{P}}^{1}(\boldsymbol{r}) - 2\pi/3$ simultaneously obtained with Figs. 4, A and B, and are strikingly similar but only when a phase-shift of $2\pi/3$ is subtracted everywhere from the measured $\Phi_{\mathrm{P}}^{1}(\boldsymbol{r})$. In Fig. 4E we show the histogram of $\delta\Phi^{1}(\boldsymbol{r}) = \Phi_{\mathrm{P}}^{1}(\boldsymbol{r}) - \Phi_{\mathrm{C}}^{1}(\boldsymbol{r})$ from Figs. 4, C, and D, and in Fig. 4F a combined histogram of all $|\delta\Phi^{i}(\boldsymbol{r})|$ $(i = 1, 2, 3)$. Hence, the relative spatial phase of the PDW and CDW states is globally $|\delta\Phi| \approx \pm 2\pi/3$. Figure 4G shows experimentally measured $N_{\mathrm{C}}^{1}(\boldsymbol{r})$ and $N_{\mathrm{P}}^{1}(\boldsymbol{r})$ merging with simultaneously measured topography $T(\boldsymbol{r})$ from the same FOV (yellow boxes in Fig. 2, E and F), revealing that an $a_0$ displacement between $N_{\mathrm{C}}^{1}(\boldsymbol{r})$ and $N_{\mathrm{P}}^{1}(\boldsymbol{r})$ generates this universal $\pm 2\pi/3$ phase shift.

**11** So, what generates and controls this complex new PDW state at atomic scale? First, Bloch-state modulations at crystal-lattice periodicity will lead inevitably to lattice-periodic modulations of $N_{\mathrm{Q}}(\boldsymbol{r})$, $N_{\mathrm{CP}}(\boldsymbol{r})$ and $\Delta(\boldsymbol{r})$ (Section 7 of (*23*)). However, at a more sophisticated and specific level, a multiband plus anisotropic energy-gap theory of $NbSe_2$ has been developed to describe superconductive electronic structure modulations at atomic scale (*36*). Beyond this, lattice-strain is important in CDW physics of TMD materials (*13,16*). Lattice-locked $3 \times 3$ commensurate CDW domains occur in $NbSe_2$, separated by discommensurations at which the CDW phase jumps by $\delta\Phi_C = \pm 2\pi/3$ (*29*). We detect these $\delta\Phi_C = \pm 2\pi/3$ discommensurations for example in $\Phi_C^1(\boldsymbol{r})$ (Fig. 4C, Fig. S11), and we find $\delta\Phi_P = \pm 2\pi/3$ phase slips for the PDW state at virtually identical locations, *i.e.* along its domain boundaries in Fig. 4D. This might be expected if the PDW is induced by the CDW coupling to the superconductivity because the PDW domains would replicate those of the preexisting CDW. Moreover, the interstate phase-difference $|\delta\Phi| = |\Phi_P(\boldsymbol{r}) - \Phi_C(\boldsymbol{r})| \approx 2\pi/3$ occurs universally (Fig. 4, C and D), not just at the commensurate domain boundaries. Hence, the simplest overall explanation is that the global phase shift $|\delta\Phi|$ does not originate from an independent lattice-lock-in of the PDW (Section 8 of (*23*)).

***12*** As to atomic-scale interactions between the CDW and the SSC states, one must consider Cooper pairing in the presence of the CDW periodic potential $V(\boldsymbol{r})$. Here, solving the linearized superconducting gap equation does generate a non-zero $\Delta_P(\boldsymbol{r})$ with $\boldsymbol{Q}_P = \boldsymbol{Q}_C$ (*37*). More intuitively, electron pairing occurs not only at momenta $(\boldsymbol{k}, -\boldsymbol{k})$, but also $(\boldsymbol{k} + \boldsymbol{G}, -\boldsymbol{k})$ and $(\boldsymbol{k}, -\boldsymbol{k} + \boldsymbol{G})$, where $\boldsymbol{G}$ is a reciprocal-lattice vector of the CDW state: $\boldsymbol{G} = m\boldsymbol{Q}_C; m = 0, \pm 1, \pm 2, \ldots$ (Fig. S12). The consequent electron-pair density at lowest order in $\boldsymbol{G}$ (Section 9 of (*23*)) is

$$\rho_P(\boldsymbol{r}) \propto \cos(\boldsymbol{Q}_C \cdot \boldsymbol{r} + \Phi_C^{\boldsymbol{Q}_C} + \delta\Phi) \tag{8}$$

Thus, the electron-pair density modulates spatially at the wavevectors $\pm\boldsymbol{Q}_C$ owing to the finite center-of-mass electron-pair momentum $(\pm\hbar\boldsymbol{Q}_C)$ imposed by the CDW. Moreover, this same approach shows that a phase difference $|\delta\Phi| = \left|\Phi_P^{\boldsymbol{Q}_C} - \Phi_C^{\boldsymbol{Q}_C}\right|$ is determined by the $\boldsymbol{k}$-space structure factor of the electron-pair wavefunction (Section 9 of (*23*)). Finally, at the

single atom scale, we find that impurity atoms leave the PDW state virtually unperturbed (Fig. S13) implying that Anderson's theorem also pertains to an *s*-wave PDW.

*13* The techniques and observations reported here herald abundant and exciting PDW physics in the many TMDs that, like $NbSe_2$, sustain both CDW and superconducting states.

**Acknowledgements:** The authors acknowledge and thank D.-H. Lee and E. Fradkin for key theoretical guidance, and M.P. Allan, K.M. Bastiaans, K. Fujita, and S.A. Kivelson for helpful discussions and advice. **Funding:** X.L. acknowledges support from Kavli Institute at Cornell. Y.X.C., R.S. and J.C.S.D. acknowledge support from the Moore Foundation's EPiQS Initiative through Grant GBMF9457. J.C.S.D. acknowledges support from Science Foundation Ireland under Award SFI 17/RP/5445 and from the European Research Council (ERC) under Award DLV-788932. **Author Contributions:** X.L, Y.X.C., and J.C.S.D. developed and carried out the experiments; X.L., Y.X.C., and R.S. developed and implemented analysis. J.C.S.D. supervised the project. The paper reflects contributions and ideas of all authors. **Competing Interests:** The authors declare no competing interests. **Data and Materials Availability:** All data are available in the main text, supplementary materials, and on Zenodo (*38*).

**Supplementary Materials:**

Materials and Methods

Supplementary Texts

Figures S1 to S16

References (*39-56*)

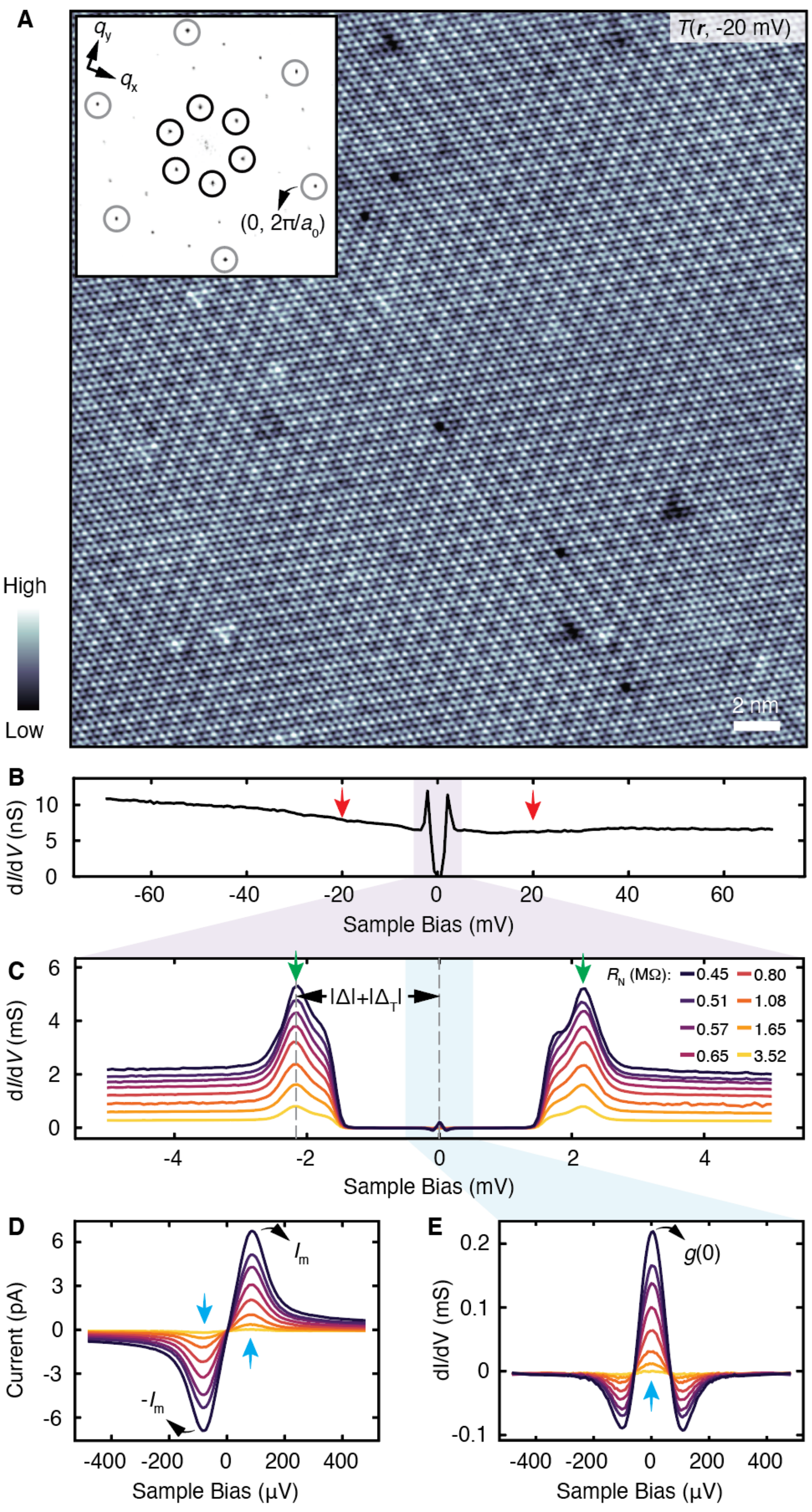

**FIG. 1 Simultaneous single-electron and electron-pair tunneling spectroscopy**

(**A**) Topographic image $T(\boldsymbol{r})$ of Se-termination surface of $NbSe_2$ measured at $T$ = 290 mK. Inset shows the Fourier transform $T(\boldsymbol{q})$ with Bragg peaks $\boldsymbol{Q}_{\mathrm{B}}^{i} \approx \{(1,0); (1/2, \sqrt{3}/2); (-1/2, \sqrt{3}/2)\} 2\pi/a_0$ indicated by grey circles, and the CDW peaks $\boldsymbol{Q}_{\mathrm{C}}^{i} \approx \{(1,0); (1/2, \sqrt{3}/2); (-1/2, \sqrt{3}/2)\} 2\pi/3a_0$ indicated by black circles.

(**B**) Typical differential tunneling conductance spectrum $g(V) \equiv dI/dV(V)$ from a Nb scan-tip to $NbSe_2$ surface at $T$ = 290 mK. The range of energies where CDW modulations are intense in $g(V)$ is indicated approximately by red arrows.

(**C**) Energy range in 1B is zoomed to show typical $g(V)$ characteristic due to the combination of the superconducting energy gaps $\Delta_{\mathrm{T}}$ of the Nb tip and $\Delta$ of the $NbSe_2$. The range of energies where superconducting coherence peaks are intense in $g(V)$ is indicated by green arrows.

(**D**) Measured electron-pair tunnel current $I_{\mathrm{CP}}(V)$ in the phase-diffusive Josephson effect energy range $|E| \lesssim 100\ \mu\mathrm{eV}$, with the range of energies where electron-pair current is maximum ($\pm I_{\mathrm{m}}$) indicated by blue arrows.

(**E**) Energy in 1C is zoomed to show phase-diffusive Josephson effect energy range, and the measured $g(V)$ whose $g(0) \propto I_{\mathrm{J}}^{2}$ from Eq. 6 indicated by a blue arrow.

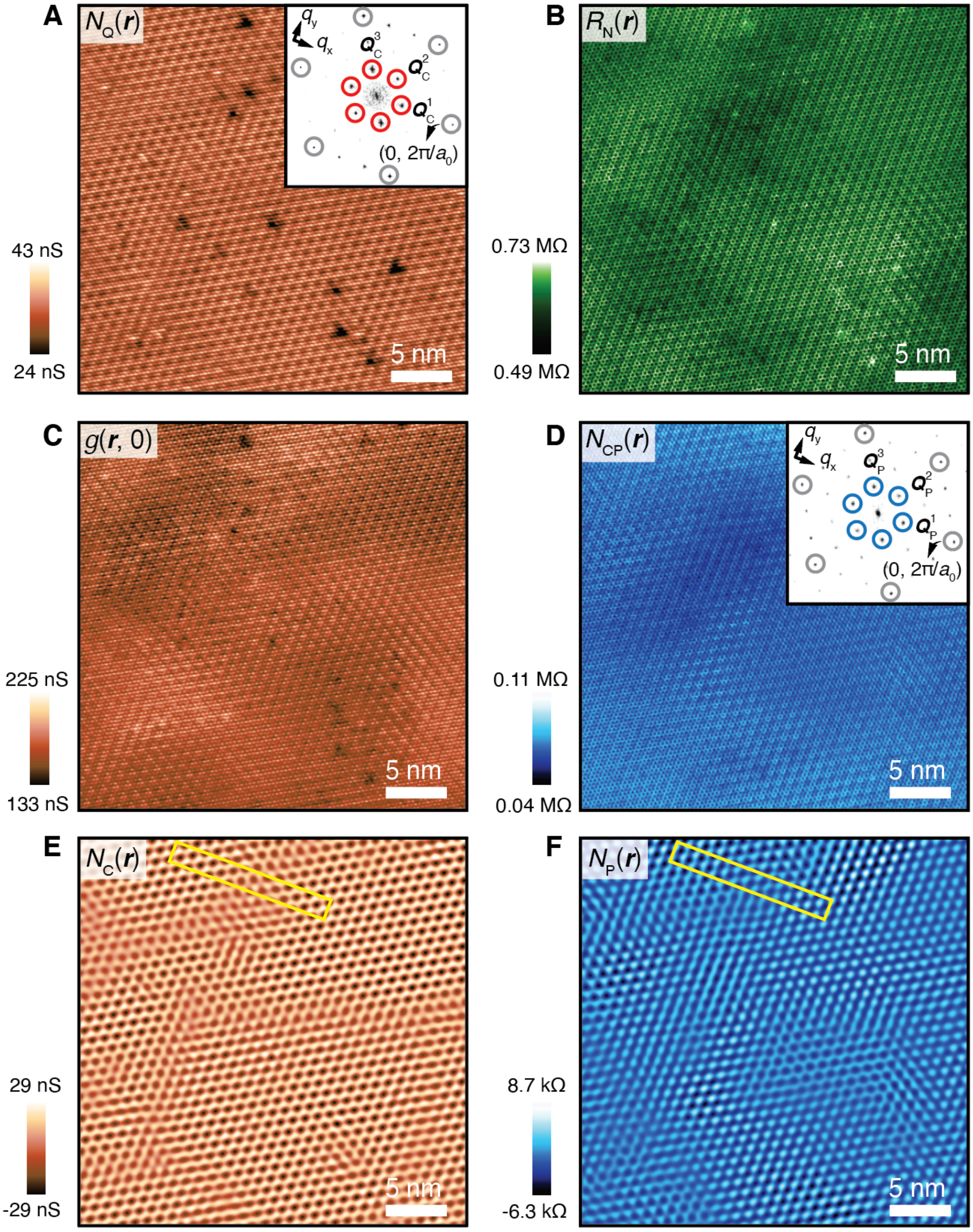

**FIG 2. Atomic-scale Electron-pair tunneling visualization of a PDW state**

(**A**) Measured $N_{\mathrm{Q}}(\boldsymbol{r}) \equiv g(\boldsymbol{r}, -20\ \mathrm{mV})$ in the same field of view (FOV) as Fig. 1A with pixel size ~30 pm at $T$ = 290 mK. Inset shows $N_{\mathrm{Q}}(\boldsymbol{q})$, with CDW peaks indicated by red circles.

(**B**) Simultaneously measured $R_{\mathrm{N}}(\boldsymbol{r}) = I^{-1}(\boldsymbol{r}, -4.5\ \mathrm{mV})$ as in (A). The purpose of this measurement is to establish the normal state tip-sample junction resistance.

(**C**) Simultaneously measured $g(\boldsymbol{r}, 0) \propto I_{\mathrm{J}}^{2}(\boldsymbol{r})$ as in (A).

(**D**) Measured Electron-pair density $N_{\mathrm{CP}}(\boldsymbol{r}) \equiv g(\boldsymbol{r}, 0) R_{\mathrm{N}}^{2}(\boldsymbol{r})$ from (B) and (C). Inset shows the PDW peaks in $N_{\mathrm{CP}}(\boldsymbol{q})$ indicated by blue circles.

(**E**) Pure CDW charge density modulations $N_{\mathrm{C}}(\boldsymbol{r})$ from (A). These are visualized at wavevectors $\boldsymbol{Q}_{\mathrm{C}}^{i} \cong \{(1,0); (1/2, \sqrt{3}/2);\ (-1/2, \sqrt{3}/2)\} 2\pi/3a_0$ by Fourier filtering $N_{\mathrm{Q}}(\boldsymbol{r})$ at the CDW peaks indicated by red circles.

(**F**) Pure electron-pair density modulations $N_{\mathrm{P}}(\boldsymbol{r})$ from (D). These are visualized at wavevectors $\boldsymbol{Q}_{\mathrm{P}}^{i} \cong \{(1,0); (1/2, \sqrt{3}/2);\ (-1/2, \sqrt{3}/2)\} 2\pi/3a_0$ by Fourier filtering $N_{\mathrm{CP}}(\boldsymbol{r})$ at the PDW peaks indicated by blue circles. Note the virtual absence of influence by impurity atoms or atomic-scale defects on the PDW state, as also seen in Fig. S13.

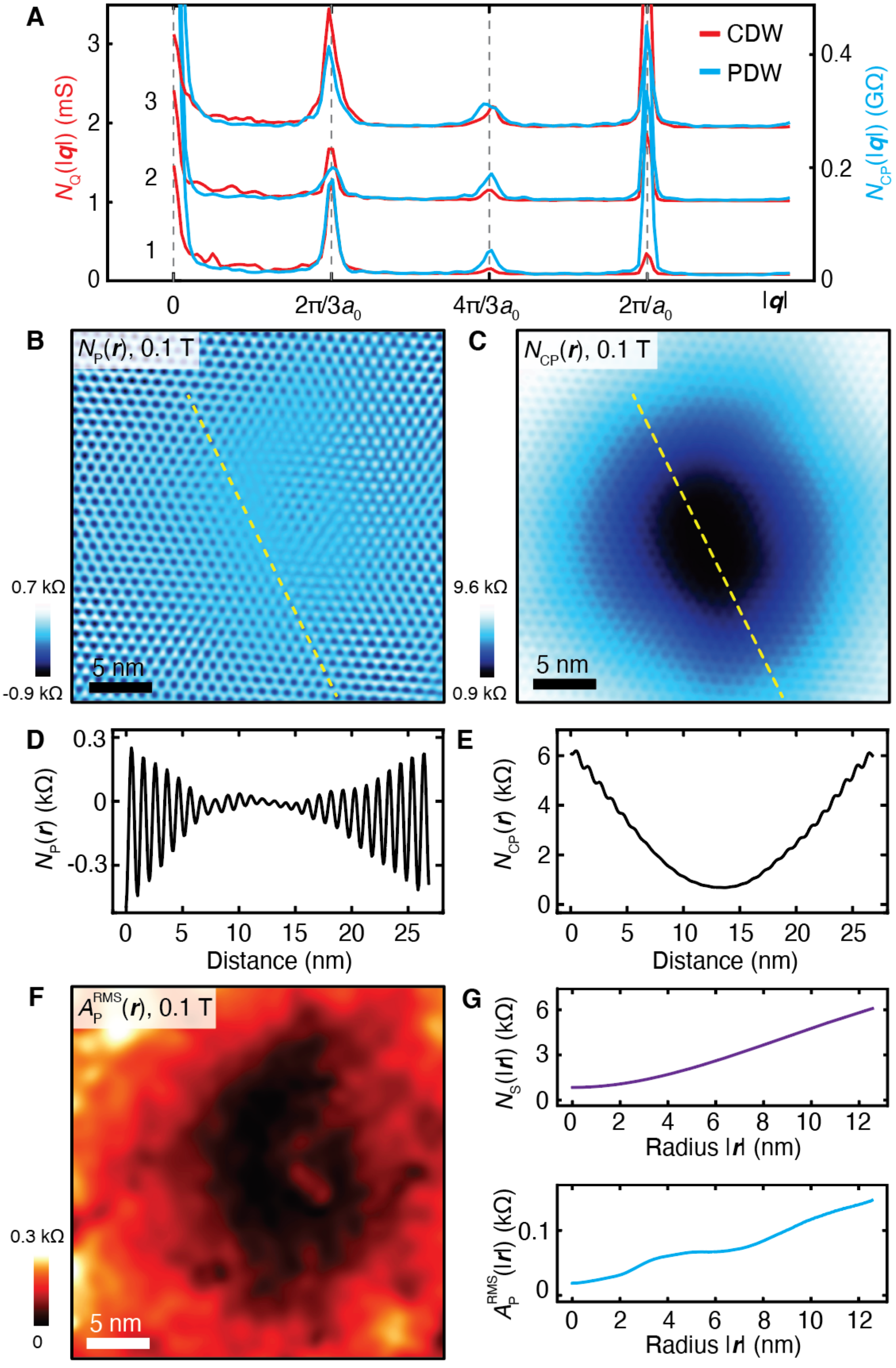

**FIG. 3 Mutual decay of superconductivity and PDW into quantized vortex core**

(**A**) Simultaneously measured amplitudes of charge density modulations $N_{\mathrm{Q}}(|\boldsymbol{q}|)$ (red) and electron-pair density modulations $N_{\mathrm{CP}}(|\boldsymbol{q}|)$ (blue) at $T$ = 290 mK, where $\boldsymbol{Q}_{\mathrm{P}}^{i} \approx \boldsymbol{Q}_{\mathrm{C}}^{i}$ is evident.

(**B**) Measured PDW electron-pair density modulations at $\boldsymbol{Q}_{\mathrm{P}}^{i}$, $N_{\mathrm{P}}(\boldsymbol{r})$, centered on the core of a quantized vortex at $B$ = 0.1 T and $T$ = 290 mK.

(**C**) Measured Electron-pair density $N_{\mathrm{CP}}(\boldsymbol{r})$ centered on vortex in (B).

(**D**) Line profile of $N_{\mathrm{P}}(\boldsymbol{r})$ along the yellow dashed line in(B).

(**E**) Line profile of $N_{\mathrm{CP}}(\boldsymbol{r})$ along the yellow dashed line in (C).

(**F**) Measured PDW amplitude $A_{\mathrm{P}}^{\mathrm{RMS}}(\boldsymbol{r})$ centered on the VC.

(**G**) The azimuthally averaged $N_{\mathrm{S}}(|\boldsymbol{r}|)$ centered on the vortex core symmetry point, and similarly the azimuthally averaged RMS amplitude of all three PDW modulations $A_{\mathrm{P}}^{\mathrm{RMS}}(|\boldsymbol{r}|)$.

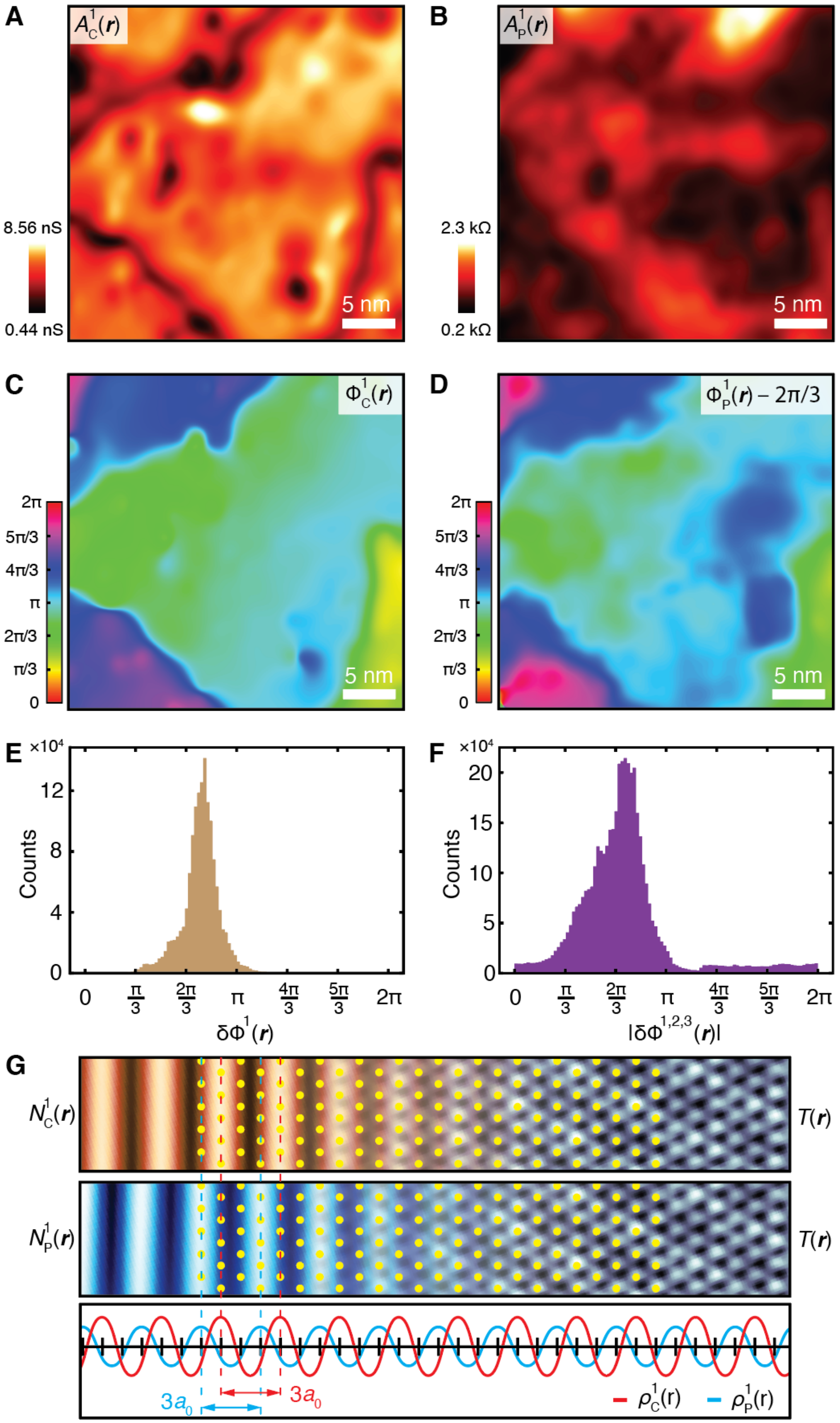

**FIG. 4 Inter-state Discommensuration of the CDW and PDW States**

(**A**) Measured CDW amplitude of $A_{\mathrm{C}}^{1}(\boldsymbol{r})$ for modulations at $\boldsymbol{Q}_{\mathrm{C}}^{1}$ from Fig. 2A. All data shown in (A)-(D) were measured in the FOV of Fig. 1A with the resulting electronic structure images shown in Fig 2.

(**B**) Simultaneously measured PDW amplitude of $A_{\mathrm{P}}^{1}(\boldsymbol{r})$ for modulations at $\boldsymbol{Q}_{\mathrm{P}}^{1}$ from 2D.

(**C**) Measured CDW spatial phase $\Phi_{\mathrm{C}}^{1}(\boldsymbol{r})$ for modulations at $\boldsymbol{Q}_{\mathrm{C}}^{1}$ (simultaneous with (A)). Our measurements find that $\Phi_{\mathrm{C}}^{1}(\boldsymbol{r})$ is virtually independent of the bias voltage $V$ for $-70mV \leq V \leq -10mV$ (Fig. S14) and impervious to atomic defects (Fig. S15).

(**D**) Measured PDW spatial phase $\Phi_{\mathrm{P}}^{1}(\boldsymbol{r}) - \frac{2\pi}{3}$ for modulations at $\boldsymbol{Q}_{\mathrm{P}}^{1}$ (simultaneous with (B)).

(**E**) Histogram of $\delta\Phi^{1}(\boldsymbol{r}) \equiv \Phi_{\mathrm{P}}^{1}(\boldsymbol{r}) - \Phi_{\mathrm{C}}^{1}(\boldsymbol{r})$ from (C),(D).

(**F**)Histograms of $|\delta\Phi^{i}(\boldsymbol{r})| \equiv |\Phi_{\mathrm{P}}^{i}(\boldsymbol{r}) - \Phi_{\mathrm{C}}^{i}(\boldsymbol{r})|$ for $i = 1, 2, 3$ from Figs. 2, A,and D. This result is also independently supported by the fact that the cross-correlation coefficient ($\eta_{\mathrm{E}} = -0.44$) between $N_{\mathrm{C}}(\boldsymbol{r})$ and $N_{\mathrm{P}}(\boldsymbol{r})$ (Figs. 2, E and F) closely matches that of simulated images ($\eta_{\mathrm{S}} \approx -0.5$) with $2\pi/3$ inter-state phase difference (Fig. S16).

(**G**) Top panels: experimentally measured $N_{\mathrm{C}}^{1}(\boldsymbol{r})$ and $N_{\mathrm{P}}^{1}(\boldsymbol{r})$ merging with simultaneously measured topography $T(\boldsymbol{r})$ from the same FOV (yellow boxes in Fig. 2, E and F) with Se atoms indicated by yellow dots. Bottom panel: schematic of the PDW:CDW inter-state discommensurations with $\delta\Phi^{1}(\boldsymbol{r}) = \frac{2\pi}{3}$.